
\newbox\leftpage \newdimen\fullhsize \newdimen\hstitle \newdimen\hsbody
\tolerance=1000\hfuzz=2pt
\def\bigans{b }
\def\answ{b }
\ifx\answ\bigans\message{(This will come out unreduced.}
\hsbody=\hsize \hstitle=\hsize 
\else\def\apans{l }\message{ lyman or hepl (l/h) (lowercase]) ? }
\read-1 to \apansw\message{(This will be reduced.}
\let\lr=L
\voffset=-.31truein\vsize=7truein
\hstitle=8truein\hsbody=4.75truein\fullhsize=10truein\hsize=\hsbody
\ifx\apansw\apans\special{ps: landscape}\hoffset=-.59truein
  \else\hoffset=.05truein\fi
\output={\ifnum\pageno=0 
  \shipout\vbox{\hbox to \fullhsize{\hfill\pagebody\hfill}}\advancepageno
  \else
  \almostshipout{\leftline{\vbox{\pagebody\makefootline}}}\advancepageno
  \fi}
\def\almostshipout#1{\if L\lr \count1=1
      \global\setbox\leftpage=#1 \global\let\lr=R
  \else \count1=2
    \shipout\vbox{\ifx\apansw\apans\special{ps: landscape}\fi 
      \hbox to\fullhsize{\box\leftpage\hfil#1}}  \global\let\lr=L\fi}
\fi
\catcode`\@=11 
\newcount\yearltd\yearltd=\year\advance\yearltd by -1900

\def\draftmode{\def\draftdate{{\rm preliminary draft:
\number\month/\number\day/\number\yearltd\ \ \hourmin}}%
\headline={\hfil\draftdate}\writelabels\baselineskip=20pt plus 2pt minus 2pt
{\count255=\time\divide\count255 by 60 \xdef\hourmin{\number\count255}
        \multiply\count255 by-60\advance\count255 by\time
   \xdef\hourmin{\hourmin:\ifnum\count255<10 0\fi\the\count255}}}

\def\nolabels{\def\eqnlabel##1{}\def\eqlabel##1{}\def\reflabel##1{}}
\def\writelabels{\def\eqnlabel##1{%
{\escapechar=` \hfill\rlap{\hskip.09in\string##1}}}%
\def\eqlabel##1{{\escapechar=` \rlap{\hskip.09in\string##1}}}%
\def\reflabel##1{\noexpand\llap{\string\string\string##1\hskip.31in}}}
\nolabels
\global\newcount\secno \global\secno=0
\global\newcount\meqno \global\meqno=1
\def\newsec#1{\global\advance\secno by1\message{(\the\secno. #1)}
\xdef\secsym{\the\secno.}\global\meqno=1
\bigbreak\bigskip
\noindent{\bf\the\secno. #1}\par\nobreak\medskip\nobreak}
\xdef\secsym{}
\def\appendix#1#2{\global\meqno=1\xdef\secsym{\hbox{#1.}}\bigbreak\bigskip
\noindent{\bf Appendix #1. #2}\par\nobreak\medskip\nobreak}
\def\eqnn#1{\xdef #1{(\secsym\the\meqno)}%
\global\advance\meqno by1\eqnlabel#1}
\def\eqna#1{\xdef #1##1{\hbox{$(\secsym\the\meqno##1)$}}%
\global\advance\meqno by1\eqnlabel{#1$\{\}$}}
\def\eqn#1#2{\xdef #1{(\secsym\the\meqno)}\global\advance\meqno by1%
$$#2\eqno#1\eqlabel#1$$}
\newskip\footskip\footskip14pt plus 1pt minus 1pt 
\def\f@@t{\baselineskip\footskip\bgroup\sevenrm\aftergroup\@foot\let\next}
\setbox\strutbox=\hbox{\vrule height9.5pt depth4.5pt width0pt}
\global\newcount\ftno \global\ftno=0
\def\foot{\global\advance\ftno by1\footnote{$^{\the\ftno}$}}
\global\newcount\refno \global\refno=1
\newwrite\rfile
\def\ref{\nref}
\def\nref#1{\xdef#1{[\the\refno]}\ifnum\refno=1\immediate
\openout\rfile=refs.tmp\fi\global\advance\refno by1\chardef\wfile=\rfile
\immediate\write\rfile{\noexpand\item{#1\ }\reflabel{#1}\pctsign}\findarg}
\def\findarg#1#{\begingroup\obeylines\newlinechar=`\^^M\pass@rg}
{\obeylines\gdef\pass@rg#1{\writ@line\relax #1^^M\hbox{}^^M}%
\gdef\writ@line#1^^M{\expandafter\toks0\expandafter{\striprel@x #1}%
\edef\next{\the\toks0}\ifx\next\em@rk\let\next=\endgroup\else\ifx\next\empty%
\else\immediate\write\wfile{\the\toks0}\fi\let\next=\writ@line\fi\next\relax}}
\def\striprel@x#1{} \def\em@rk{\hbox{}} {\catcode`\%=12\xdef\pctsign{
\def\semi{;\hfil\break}
\def\addref#1{\immediate\write\rfile{\noexpand\item{}#1}} 
\def\listrefs{\immediate\closeout\rfile
\baselineskip=14pt\centerline{{\bf References}}\bigskip{\frenchspacing%
\escapechar=` \input refs.tmp\vfill\eject}\nonfrenchspacing}
\def\startrefs#1{\immediate\openout\rfile=refs.tmp\refno=#1}
\def\figures{\centerline{{\bf Figure Captions}}\medskip\parindent=40pt}
\def\fig#1#2{\medskip\item{Fig.~#1:  }#2}
\catcode`\@=12 
\def\tr{{\rm tr}} \def\Tr{{\rm Tr}}
\def\lspace{\ifx\answ\bigans{}\else\qquad\fi}
\def\lbspace{\ifx\answ\bigans{}\else\hskip-.2in\fi} 
\def\boxeqn#1{\vcenter{\vbox{\hrule\hbox{\vrule\kern3pt\vbox{\kern3pt
        \hbox{${\displaystyle #1}$}\kern3pt}\kern3pt\vrule}\hrule}}}
\def\mbox#1#2{\vcenter{\hrule \hbox{\vrule height#2in
                \kern#1in \vrule} \hrule}}  
\magnification 1200
\def\IR{\relax{\rm I\kern-.18em R}}
\def\IB{\relax{\rm I\kern-.18em B}}
\def\overcirc#1{\ontopof{#1}{\circ}{1.2}\mathord{\box2}}
\def\ontopof#1#2#3{%
{\mathchoice
{\oontopof{#1}{#2}{#3}\displaystyle\scriptstyle}%
{\oontopof{#1}{#2}{#3}\textstyle\scriptstyle}%
{\oontopof{#1}{#2}{#3}\scriptstyle\scriptscriptstyle}%
{\oontopof{#1}{#2}{#3}\scriptscriptstyle\scriptscriptstyle}}}
\def\oontopof#1#2#3#4#5{%
\setbox0=\hbox{$#4#1$}%
\setbox1=\hbox{$#5#2$}%
\setbox2=\hbox{}\ht2=\ht0 \dp2=\dp0 %
\ifdim\wd0>\wd1 %
\setbox1=\hbox to\wd0{\hss\box1\hss}%
\mathord{\rlap{\raise#3\ht0\box1}\box0}%
\else   %
\setbox1=\hbox to.9\wd1{\hss\box1\hss}%
\setbox0=\hbox to\wd1{\hss$#4\relax#1$\hss}%
\mathord{\rlap{\copy0}\raise#3\ht0\box1}%
\fi}

\ref\gold{J.  Goldstone and R. Jackiw, {\it Phys.  Lett.} {\bf 74B}
(1978) 81.} \ref\simonov{Y.A.  Simonov, {\it Sov.  J. Nucl.  Phys.} {\bf
41} (1985) 835.} \ref\ken{K.  Johnson, in {\it QCD -- 20 Years Later},
Aachen, June 1992.} \ref\badalyan{A.M.  Badalyan, {\it Sov.  J. Nucl.
Phys.} {\bf 38} (1983) 464.} \ref\wuyang{T.T.  Wu and C.N.  Yang, {\it
Phys.  Rev.} {\bf D12} (1975) 3845.} \ref\luscher{M.  L\"uscher and G.
M\"unster, {\it Nucl.  Phys.} {\bf B232} (1984) 445.} \ref\baal{J.
Koller and P. van Baal, {\it Nucl.  Phys.} {\bf B302} (1988) 1.}
\ref\straumann{N.  Straumann, {\it General Relativity and Astrophysics},
Springer-Verlag, 1984.} \ref\deser{S.  Deser and F. Wilczek, {\it Phys.
Lett.} {\bf 65B} (1976) 391.} \eject

\def\footnoterule{\kern-3pt \hrule width \hsize \kern2.6pt}
\pageno=0
\footline={\ifnum\pageno>0 \hss\folio\hss \else\fi}

\centerline{{\bf THE HIDDEN SPATIAL GEOMETRY OF NON-ABELIAN GAUGE
THEORIES}  \footnote{*} {\sevenrm This work is supported in part by
funds provided by the U. S. Department of Energy (DOE) under contract
\#DE-AC02-76ER03069, by NSF Grant PHY/9206867, AEN 90-0033 Grant
(Spain), NATO Grant \#910890, and by M.E.C.  (Spain). \hfil\break
\hbox{~~~}$^{\rm (a)}$~Permanent address: Department of Mathematics and
Center for Theoretical Physics, Massachusetts Institute of Technology,
Cambridge MA~~02139--4307.}}
\vskip15pt
\centerline{\tenrm Daniel~Z.~Freedman$^{\rm (a)}$}
\vskip-2pt
\centerline{\sevenrm  Theory Division, CERN, CH--1211, Geneva 23,
Switzerland} \vskip6pt
\centerline{\tenrm Peter~E.~Haagensen}
\vskip-2pt
\centerline{\sevenrm  Department d'Estructura i Constituents de la
Mat\`eria, Universitat de Barcelona, 08028 Barcelona, Spain}
\vskip6pt
\centerline{\tenrm Kenneth~Johnson}
\vskip-2pt
\centerline{\sevenrm  Center for Theoretical Physics, Laboratory for
Nuclear Science and Department of Physics,}
\vskip-4pt
\centerline{\sevenrm Massachusetts Institute of Technology,
Cambridge, MA 02139--4307, U.S.A.}
\vskip6pt
\centerline{\tenrm Jos\'e~I.~Latorre}
\vskip-2pt
\centerline{\sevenrm  Department d'Estructura i Constituents de la
Mat\`eria, Universitat de Barcelona, 08028 Barcelona, Spain}\vskip10pt

\vfill

\centerline{\bf ABSTRACT}
\midinsert
\baselineskip=13pt plus 1pt
\smallskip
The Gauss law constraint in the Hamiltonian form of the $SU(2)$ gauge
theory of gluons is satisfied by any functional of the gauge invariant
tensor variable $\phi^{ij} = B^{ia} B^{ja}$.  Arguments are given that
the tensor $G_{ij} = (\phi^{-1})_{ij}\,\det B$ is a more appropriate
variable.  When the Hamiltonian is expressed in terms of $\phi$ or $G$,
the quantity $\Gamma^i_{jk}$ appears.  The gauge field Bianchi and Ricci
identities yield a set of partial differential equations for $\Gamma$ in
terms of $G$.  One can show that $\Gamma$ is a metric-compatible
connection for $G$ with torsion, and that the curvature tensor of
$\Gamma$ is that of an Einstein space.  A curious 3-dimensional spatial
geometry thus underlies the gauge-invariant configuration space of the
theory, although the Hamiltonian is not invariant under spatial
coordinate transformations.  Spatial derivative terms in the energy
density are singular when $\det G=\det B=0$.  These singularities are
the analogue of the centrifugal barrier of quantum mechanics, and
physical wave-functionals are forced to vanish in a certain manner near
$\det B=0$.  It is argued that such barriers are an inevitable result of
the projection on the gauge-invariant subspace of the Hilbert space, and
that the barriers are a conspicuous way in which non-abelian gauge
theories differ from scalar field theories.
\endinsert
\vfill
\centerline{Submitted to: {\it Nuclear Physics B}}
\vfill
\line{CTP \#2238  \hfil August 1993}
\eject

\goodbreak\bigskip
\noindent{\bf 1.\quad Introduction}
\medskip\nobreak

The implementation of the Gauss law constraint on physical states in the
Hamiltonian form of non-abelian gauge theories is a major obstacle for
non-perturbative studies.  Since the difficulty in treating Gauss' law
stems from the non-covariant gauge transformation of the vector
potential $A^ a_i$, one can attempt to solve the problem by formulating
the theory in terms of variables which transform covariantly.  For
example, Goldstone and Jackiw \gold\ suggested the use of the electric
field as the fundamental variable.  Their approach led to an exact
implementation of Gauss' law in the $SU(2)$ gauge theory of gluons, but
to a complicated Hamiltonian which has not to our knowledge been used in
concrete calculations.

One may also consider the problem in magnetic variables.  Here Simonov
\simonov\ has applied a ``polar representation" of the vector potential
which allows the removal of gauge degrees of freedom, but gives a
Hamiltonian and functional measure which are non-local.  A year ago, one
of us \ken\ proposed a canonical transformation to the magnetic field
$B^{ia}$ and a conjugate $C^a_i$, both of which tranform homogeneously.
The Gauss law constraint is then satisfied by state functionals
$\psi[\phi^{ij}]$ which depend on the gauge invariant tensor variable
$\phi^{ij}=B^{ia}B^{ja}$.  (See \badalyan\ for an earlier, less complete
proposal to use the magnetic field).

In this paper we review briefly the proposal of \ken\ and then discuss
further ideas related to the use of gauge-invariant magnetic variables
in the $SU(2)$ gluon theory.  For most of the considerations it is not
necessary to make a complete canonical transformation.  Rather we
observe that even if $A^a_i$ is taken as the fundamental variable,
states $\psi[\phi^{ij}]$ satisfy the Gauss law constraint, and it is
sensible to study the form of the Hamiltonian for such states.  There
are several reasons why the variable $G_{ij}=\phi_{ij} \det B$, where
$\phi_{ik}\phi^{kj}=\delta_i^j$, is more appropriate than $\phi^{ij}$,
and we also obtain the form of the Hamiltonian for functionals
$\psi[G_{ij}]$.

In both cases a gauge invariant quantity $\Gamma^i_{jk}$ appears in the
electric energy density, and the gauge field Bianchi and Ricci
identities lead to partial differential equations from which one can, in
principle, determine $\Gamma^i_{jk}$ in terms of $G_{ij}$.  An
unexpected geometrical structure then emerges.  If $G_{ij}$ is viewed as
an (indefinite) metric tensor on ${\IR}^3$, then $\Gamma^i_{jk}$ is a
metric-compatible affine connection with torsion, and the Riemann
${R^i}_{jk\ell}$ and Ricci $R_{ij}$ tensors of $\Gamma$ are those of a
3-dimensional Einstein space.  The Hamiltonian is always expressed in
Cartesian coordinates on ${\IR}^3$.  It cannot be and is not
diffeomorphism invariant because it involves the Cartesian metric
$\delta_{ij}$ as well as $G_{ij}$.  It turns out that the energy density
transforms in a specific tensor representation of $GL(3)$.

It can be argued that $\phi^{ij}$ or, somewhat more precisely, $G_{ij}$
are symmetric tensors whose six independent components describe the
local gauge invariant degrees of freedom of the $SU(2)$ gauge theory.
It is then curious that the gauge invariant phase space admits a fairly
natural spatial geometry while the Hamiltonian is not invariant.  Thus,
two configurations of $G_{ij}$ mathematically related by a
``diffeomorphism" describe physically distinct gauge field
configurations, typically with different energies.

To clarify things it should be stated that given $A^a_i(x)$, one can
calculate $B^{ia}(x)$, $G_{ij}(x)$ and $\Gamma^k_{ij}(x)$ by
straightforward local formulas, and it then turns out that $G_{ij}(x)$
and $\Gamma^k_{ij}(x)$ are the metric and connection of a 3-dimensional
Einstein geometry with torsion.  We refer to this as the forward map.
It is then an interesting question whether this transformation can be
inverted.  For $SU(2)$ it is easy to show that given $G_{ij}(x)$ and
$\Gamma^k_{ij}(x)$, one can reconstruct $B^{ia}(x)$ and $A^a_{i}(x)$, up
to a gauge transformation, by a local construction.  Reasonable but
non-rigorous arguments are given that for a given configuration
$G_{ij}(x)$ the Einstein space condition $$
R_{ij}(\Gamma)=-2G_{ij}\eqno{(1.1)} $$ can be solved to obtain the
contortion tensor ${K_{ij}}^k$ which is related to $\Gamma$ by $$
\Gamma^i_{jk}=\overcirc{\Gamma}^{i}_{jk}-{K_{jk}}^i\eqno{(1.2)} $$ where
$\overcirc{\Gamma}$ is the standard Christoffel symbol.  In general,
${K_{ij}}^k$ is nonlocally related to $G_{ij}$, and we expect two
solutions for ${K_{ij}}^k$ for each configuration of $G_{ij}$.  Thus the
geometry appears to generate a double-valued map from $G_{ij}$ to a
magnetic field $B^{ia}$, unique up to a gauge, and a pair of potentials
$A^a_i$.  This structure seems to be consistent with the Wu-Yang
ambiguity \wuyang . Some examples of specific geometries and the gauge
field configurations related to them are studied.

Our motivation in formulating the theory in terms of gauge-invariant
local variables was to implement Gauss' law exactly, so that the
resulting Hamiltonian in the physical subspace could be used for
dynamical calculations of the vacuum structure and glueball spectrum.
It is the slowly-varying modes (compared to ${1\over \Lambda_{\rm
QCD}}$) of the system for which non-perturbative treatment is most
urgently required, and the use of gauge-invariant variables allows a
gauge-invariant definition of slow variation.  It is not fully clear how
to implement dynamical calculations or whether the geometrical structure
discussed above will be useful for this purpose.

Nevertheless, there are some physical implications because the energy
density is singular when $\det \phi^{ij}=\left(\det G_{ij}\right)^2=
\left( \det B^{ia} \right)^2=0$.  These conditions correspond to
coordinate singularities of the gauge-invariant configuration space and
a singularity of the transformation $ A^a_i \longrightarrow B^{ia} $,
just as $ r=0 $ is a singular point of the spherical coordinate system
and of the transformation from Cartesian coordinates.  Thus, one can
interpret the singularities as the gauge theory analogue of the
centrifugal barrier of quantum mechanics.  Such barriers also occur in
the electric formulation \gold\ of non-abelian gauge theory and appear
to be an inevitable result of the ``projection" onto the gauge-invariant
physical subspace.  There are no such barriers in scalar field theories.
Finite energy eigenfunctionals or variational trial functionals must
satisfy certain vanishing conditions near the singularity, and these
conditions involve a complicated combination of functional and spatial
derivatives.  It is then suggested that a better understanding of these
barriers may provide qualitative insight into the dynamics of
non-abelian gluons.

\goodbreak\bigskip
\noindent{\bf 2.\quad  Gauss' Law and the Variables $\phi^{ij}$ and
$G_{ij}$} \medskip\nobreak

We begin with the $SU(2)$  Yang-Mills Hamiltonian in the $A_0^a(x)=0$
gauge:
$$
H = {1\over2} \int d^3 x \left[ g^2 \left(E^{ia}(x)\right)^2 +
{1\over g^2} \left( B^{ia}(x) \right)^2 \right] ~~,
\eqno{(2.1)}
$$
where the electric and magnetic fields are
$$
E^{ia} \equiv {1\over g^2} \dot{A}_i^a ~, ~~~~~~~
B^{ia} \equiv \epsilon^{ijk}
\left(\partial_j A^a_k+{1\over2}\epsilon^{abc} A_j^b \, A_k^c \right)
{}~.
\eqno{(2.2)}
$$
The standard pair of canonical variables are the gauge potential and the
electric field, with equal-time commutators:
$$
\left[ A_i^a(x), E^{jb}(y) \right]
= i \delta^{ab} \delta_{i}^{j} \, \delta^{(3)} (x-y) ~.
\eqno{(2.3)}
$$

The generator of the gauge transformation with parameter $\theta^a(x)$
is
$$
\eqalign{
{\cal G}[\theta]&=\int d^3\!x \,\theta^a (x) \, {\cal G}^a(x) \cr
{\cal G}^a(x) &= D_k \, E^{ka}(x)
= \partial_k E^{ka} + \epsilon^{abc} A_k^b E^{kc} ~,}
\eqno{(2.4)}
$$
and the quantum transformation  rules of the local fields and the
Hamiltonian are:
$$
\eqalign{
\delta A_i^a(x) &= i \left[ {\cal G}[\theta], A_i^a(x) \right] =
D_i \theta^a (x) =
\partial_i \theta^a (x) +\epsilon^{abc} A_i^b(x) \, \theta^c(x) \cr
\delta E^{ia}(x) &= i \left[ {\cal G}[\theta], E^{ia}(x) \right] =
\epsilon^{abc} E^{ib}(x) \, \theta^c(x) \cr
\delta B^{ia}(x) &= i \left[ {\cal G}[\theta], B^{ia}(x) \right] =
\epsilon^{abc} B^{ib}(x) \, \theta^c(x) \cr
&\,\,\, \left[ {\cal G} [\theta],~H \right] = 0 ~.
}
\eqno{(2.5)}
$$
Gauss' law, ${\cal G}^a(x)=0$, is one of the classical equations of motion
obtained from the Lagrangian of the theory before gauge fixing.  In the
gauge-fixed quantum theory one must impose it as the constraint
$$
{\cal G}^a(x) \, | \, \psi_{\rm phys} \, \rangle = 0
\eqno{(2.6)}
$$
on physical states in the Hilbert space.  It is the implementation of this
constraint that motivates the transformation \ken\ which we now discuss.

Let us consider the possibility of describing the configuration space of
the system using $B^{ia}(x)$ rather than $A_i^a(x)$.  In three spatial
dimensions (and only three), they have the same number of spatial
components, but this is certainly not the only consideration.  The
magnetic field satisfies the Bianchi identity

$$
\partial_i B^i(x) = 0 \eqno{(2.7)}
$$
in the abelian case, and
$$
D_iB^{ia}=\partial_iB^{ia}(x)+\epsilon^{abc}A_i^b(x)\,B^{ic}(x)=0
\eqno{(2.8)} $$
for the non-abelian theory.  Thus, $B^i$ is constrained
to only two independent components in the abelian case, and cannot be
used as a variable.  The non-abelian Bianchi identity, on the other
hand, is not a constraint on $B^{ia}$, but rather a relation between
$B^{ia}$ and $A_i^a$, which is compatible with (2.2), and so presents no
immediate obstruction to our goal.

In \ken\ a formal canonical transformation from the conjugate variables
$(A_i^a,~E^{ja})$ to a new set $(B^{ia}, C^a_j)$ was presented.  Now we
take the simpler viewpoint that $B^{ia}$ is a useful, dependent variable
on the original configuration space.  The action of the electric field
on functionals $\psi[B]$ is determined by the chain rule
$$
\eqalign{
E^{ia} = -i {\delta \over \delta A_i^a(x)}
&= -i \int d^3\!y \, {\delta B^{jb}(y)\over\delta A_{i}^{a}(x)}
\, {\delta \over \delta B^{jb}(y)} \cr
&= - i \epsilon^{ijk} \, D_j \, {\delta \over \delta B^{ka}} ~~.}
\eqno{(2.9)}
$$
The generator of gauge transformations then acts as
$$\eqalign{
{\cal G}^a(x) \, \psi[B] &= D_i E^{ia} \psi[B] \cr
&= -i \epsilon^{ijk} D_i D_j {\delta \over \delta B^{ka}} \psi[B] \cr
&= -i \epsilon^{abc} B^{ib}(x) {\delta \over \delta B^{ic}(x)}
\psi[B]\, .\cr}
\eqno{(2.10)}
$$
This is of the form of an ``angular momentum'' because $B$ and its
canonical conjugate $C=-i \, \delta / \delta B$ transform homogeneously.

Next we note that the positive symmetric tensor $\phi^{ij}=B^{ia}
B^{ja}$ is gauge invariant, so that any functional $\psi[\phi]$
satisfies the physical state constraint (2.6).  Although $\phi^{ij}$ has
6 independent components, which is the correct number necessary to
describe the gauge invariant content of a field configuration, it is not
quite satisfactory because it does not give a complete description of
the gauge-invariant subspace of the configuration space.  Naturally,
$\det B$ is an independent invariant, but $(\det B)^2$ can be expressed
in terms of $\phi$, so only the sign of $\det B$ is independent.  Thus,
to use $\phi$ one must introduce a discrete label $\alpha=\pm$ and
consider $\psi[\phi,\alpha]$.

Alternatively we avoid this problem by introducing another gauge
invariant variable which has the same dimension as $B^{ia}$, namely the
tensor $$G_{ij} = {B^a}_{i} {B^a}_{j}\det B ~, \eqno{(2.11)}$$ where
$B^{ia}{B^a}_{j}= \delta_j^i$.  One finds that $\det G = \det B$, and
that $G_{ij}$ is either positive- or negative-definite.  The relation
between $G$ and $\phi$ is
$$
\phi^{ij}={1\over2}\epsilon^{ik\ell}\epsilon^{jmn}G_{km}G_{\ell n}
= G^{ij} \det G ~, \eqno{(2.12)}
$$
and one may show that
$$
F^a_{\,ij}F^a_{\,k\ell}=G_{ik}G_{j\ell}-G_{jk} G_{i\ell} ~.
\eqno{(2.13)}
$$
The magnetic energy density is
$$
{1\over2g^2} \, \delta_{i\bar{\imath}} \, B^{ia}
\, B^{\bar{\imath} a}\, |\psi |^2
= {1\over2g^2}\,\delta_{i\bar{\imath}}\,\phi^{i\bar{\imath}}\,|\psi |^2
= {1\over2g^2}(\delta^{j\bar{\jmath}}\delta^{k\bar{k}}-\delta^{j\bar{k}}
\delta^{k \bar{\jmath}})G_{j\bar{\jmath}}G_{k\bar{k}} \, |\psi |^2
\eqno{(2.14)}
$$
The variable $G_{ij}$ was used previously to describe constant gauge
fields on the torus by L\"uscher \luscher\ and others \baal . Indeed for
constant potentials $A^a_{\,i}$, (2.11) reduces to
$$
G_{ij} = A^a_{\,i} A^a_{\,j} \eqno{(2.15)}
$$
which is exactly the variable used in \luscher .

\goodbreak\bigskip
\noindent{\bf 3.\quad Geometry}
\medskip\nobreak

The plan now is to work out the form taken by the electric energy
density in gauge-invariant variables.  As an intermediate step we use
the $\phi$ variable, but later return to $G$.  The functional chain rule
gives
$$
\eqalign{
{\delta \over \delta A^a_{\,i}} \psi [\phi]
&= {\delta B^{kb} \over \delta A^a_{\,i}}
{\delta \phi^{mn} \over \delta B^{kb}}
{\delta \psi [\phi] \over \delta \phi^{mn}}\cr
&= 2 \epsilon^{ijk} D_j^{\,ab}
\left( B^{\ell b}{\delta\psi\over\delta\phi^{k\ell}} \right) \cr
&= 2 \epsilon^{ijk}
\left[ B^{\ell a} {\partial_j}
{\delta \psi \over \delta \phi^{k\ell}}
+D_j B^{\ell a} {\delta \psi \over \delta \phi^{k\ell}} \right] ~~.
}\eqno{(3.1)}
$$
We define a connection-like quantity $\Gamma'^{m}_{j\ell}$ by
$$
B^{\ell a} \, \Gamma'^{m}_{j\ell} \equiv - D_j \, B^{ma}
\eqno{(3.2)}
$$
so that (3.1) can be rewritten as
$${\delta \over \delta A^a_i} \psi[\phi]
= 2 \epsilon^{ijk} B^{\ell a} \left[
\delta_\ell^{\,m}{\partial_j}  - \Gamma'^{m}_{j\ell} \right]
{\delta \psi \over \delta\phi^{km}}\, .
\eqno{(3.3)}$$

Let us analyze the properties of $\Gamma'$, first multiplying (3.2) by
$B^{ia}$ to get
$$
\phi^{i\ell} \Gamma'^{m}_{j\ell}
\equiv - B^{ia} D_j B^{ma} ~~. \eqno{(3.4)}
$$
The $im$ symmetric part of this equation can be written as
$$
\partial_j \phi^{im}
+ \Gamma'^{i}_{j\ell} \phi^{\ell m}
+ \Gamma'^{m}_{j\ell} \phi^{i\ell}
= 0
\eqno{(3.5)}
$$
which looks like a metric-compatible condition for the (inverse) metric
$\phi^{im}$.
Next differentiate (3.2) and manipulate as follows
$$
\eqalign{
D_i \left( B^{a\ell} \Gamma'^{m}_{j\ell} \right) &\equiv
- D_i D_j B^{am} \cr
B^{ak}
\left(
\partial_i \Gamma'^{m}_{jk}- \Gamma'^{\ell}_{ik}\Gamma'^{m}_{j\ell}
\right)
&= -D_i D_j B^{am}} \eqno{(3.6)}
$$

With the help of the gauge theory Ricci identity,
the $ij$ anti-symmetric part of (3.6) can be written as
$$
\eqalign{
\partial_i \Gamma'^{m}_{jk}
- \partial_j \Gamma'^{m}_{ik}
- \Gamma'^{\ell}_{ik} \Gamma'^{m}_{j\ell}
+ \Gamma'^{\ell}_{jk} \Gamma'^{m}_{i\ell}
&=
-{B^{a}}_{k} [D_i,~D_j] B^{ma}\cr
&=-\epsilon^{abc}\epsilon_{ij\ell}{B^{a}}_{k}B^{\ell b} B^{mc} \cr
&=-\epsilon_{ij\ell}\epsilon^{\ell mn}{B^{a}}_{k}{B^{a}}_{n}\det B\cr
&= \delta_j^{\,m} G_{ik} - \delta_i^{\,m} G_{jk}}\eqno{(3.7)}
$$
A geometrical interpretation is not yet at hand, because (2.12) shows
that both $\phi$ and $G$ cannot transform as tensors, which would be
necessary for (3.5) and the curvature-like (3.7) to be compatible.
Since (3.7) signals that $G$ is tensorial, we insert (2.12) in (3.5),
and substitute
$$
\Gamma'^{i}_{j\ell}=-{1\over2}\delta_\ell^i\,\partial_j\ln\det G +
\Gamma_{j\ell}^i ~~, \eqno{(3.8)}
$$
or equivalently,
$$
{B^{\ell a}\over \sqrt{\det G}}\Gamma^i_{j\ell}
= -D_j\left( {B^{ia}\over \sqrt{\det G}}\right)\, .\eqno{(3.9)}
$$
Then, (3.5) becomes
$$
\partial_j G^{im} + \Gamma^i_{j\ell} G^{\ell m}
+ \Gamma^{m}_{j\ell} G^{i\ell} = 0 ~~.
\eqno{(3.10)}
$$
The $\partial_j\ln\det G$ term drops out on the left side of (3.7), and
we find that the curvature tensor of the $\Gamma$ connection satisfies
$$
{R^\ell}_{kij}\equiv\partial_i\Gamma^\ell_{jk}
-\partial_j\Gamma^\ell_{ik}-\Gamma^m_{ik}
\Gamma^\ell_{jm}+\Gamma^m_{jk}\Gamma^\ell_{im}
= \delta^\ell_jG_{ik}-\delta^\ell_iG_{jk}\, ,\eqno{(3.11)}
$$
while the Ricci tensor obeys
$$
R_{kj}=-2G_{kj}\, .\eqno{(3.12)}
$$
One more contraction gives $R=-6$ for the Ricci scalar.
One may now interpret (3.10) as a metric compatibility condition for
$\Gamma$ with respect to $G$, so that $\Gamma$ must take the form of a
connection with torsion, namely
$$\eqalign{
\Gamma^i_{jk}&=\overcirc{\Gamma}^{i}_{jk}-{K_{jk}}^i\cr
\overcirc{\Gamma}^{i}_{jk}&={1\over 2}G^{i\ell}
(\partial_jG_{\ell k}+\partial_kG_{j\ell}-\partial_\ell G_{jk})\cr
K_{jki}&=-K_{jik}\, .}
\eqno{(3.13)}$$
Clearly (3.12) is the Einstein condition for the
(in principle non-symmetric) Ricci tensor.

For a metric-compatible connection,
$$
R_{\ell kij}=-R_{k\ell ij}\eqno{(3.14)}
$$
in addition to the manifest antisymmetry in $ij$. One can then easily show
(using $\epsilon$-tricks) that, even with torsion, the Riemann tensor for
$D=3$ is fully determined by its contractions and takes the form
$$
R_{ijk\ell}=G_{ik}R_{j\ell}- G_{i\ell}R_{jk}- G_{jk}R_{i\ell}
+ G_{j\ell}R_{ik}- {R\over2}( G_{ik}G_{j\ell}- G_{i\ell}G_{jk}) ~.
\eqno{(3.15)}
$$
Thus (3.11) and the simpler (3.12) have the same content.

An  apparent further constraint on the geometry follows from the gauge
theory Bianchi identity (2.8). When applied to (3.4) one finds that
$\Gamma^{'i}_{ik}= 0$. Using (3.8) and (3.13) we see that this is
equivalent to the additional trace condition on the contortion
$$
{K^j}_{jk}=0\eqno{(3.16)}
$$
which implies that $K$ can be represented as
$$
{K^i}_{jk}=\epsilon_{jkn}S^{ni}{1\over\sqrt{\det G}}\, ,\eqno{(3.17)}
$$
where $S^{ni}$ is a symmetric tensor and $\epsilon_{jkn}$ is defined
at the end of this Section.

It turns out that (3.16) can also be derived directly from geometry
without reference to gauge theory.  To do this one starts with a
contracted form of the second Bianchi identity of a curvature tensor
with torsion \straumann , and uses (3.11) and (3.12) to replace the
Riemann and Ricci tensors by metrics.  This quickly gives (3.16) which
can thus be viewed as an integrability condition for an Einstein space
with torsion.

Equations (3.10 -- .17) completely define the spatial geometry
associated with the gauge-invariant subspace of the configuration space
of $SU(2)$ gauge theory.  It is worthwhile to emphasize that given a
potential $A_i^a$, the magnetic field $B^{ia}$, the metric $G_{ij}$, and
the connection $\Gamma^i_{jk}$ can be calculated directly from the
formulas (2.2), (2.11), and (3.9) respectively. $K$ and $R$ can then be
calculated through (3.13), and all geometrical conditions are then
satisfied.  Later we will begin to address the converse question, namely
given a symmetric tensor $G_{ij}$, can one find a contortion tensor
$K_{ijk}$ satisfying (3.16) and such that the Einstein condition (3.12)
is satisfied.  This is essentially the question whether the change of
variables $A_i^a\rightarrow B^{ia}\rightarrow G_{ij}$ is invertible.

We now need some notation to cope with the original fiber-bundle
geometry of the gauge theory and the new spatial geometry.  We use $D_i$
to denote the gauge theory covariant derivative, as implicitly defined
in (2.4 -- .5), which ``sees" only gauge indices.  Then $\nabla_i$ and
$\overcirc{\nabla}_i$ are used to denote spatial derivatives with and
without torsion, e.g., on a covariant vector
$$\eqalign{
\nabla_iV_j&\equiv\partial_iV_j-\Gamma^k_{ij}V_k\cr
\overcirc{\nabla}_i V_j&\equiv\partial_iV_j-\overcirc{\Gamma}^{k}_{ij}
V_k ~.} \eqno{(3.18)}
$$
Later we will use $R$ and $\overcirc{R}$ to denote curvatures with and
without torsion.  The Levi-Civita density $\epsilon^{ijk}$ takes the
usual values $\pm 1,0$, and one applies 3 factors of $G_{i\ell}$, etc.
to obtain $\epsilon_{\ell mn}$ so that $\epsilon_{\ell mn}/\sqrt{\det
G}$ is a tensor.  Indeed, indices of all quantities are raised and
lowered from their initially defined form with $G$, while all
contractions with the Cartesian metric $\delta_{ij}$ are indicated
explicitly.  (The exceptions to these conventions are that $B^{ia}$,
${B^a}_i$ and $\phi^{ij}$, $\phi_{ij}$ are matrix inverses, and
$\epsilon_{ijk}$ in (3.7) $=\{ \pm 1,0\}$.  Throughout, $\sqrt{\det G}$
means $\sqrt{|\det G|}$.)

\goodbreak\bigskip
\noindent{\bf 4.\quad The Energy Density in Geometric Variables}
\medskip\nobreak

Let us return to a more physical question, namely, the form of the
Hamiltonian.  With gravity neglected, this cannot be invariant under
spatial diffeomorphisms.  This is already clear from the magnetic energy
density (2.14) which involves contractions of the true metric
$\delta_{ij}$ and the gauge-invariant tensor $G_{ij}$.

For the electric energy density we might expect noncovariance both due
to the $\delta_{ij}$ contraction, and because the derivative appearing
in (3.3) does not appear to be fully covariant.  It turns out, however,
that this is not the case, and in fact we find that the {\it unique}
source of noncovariance in the energy density lies in the $\delta_{ij}$
contraction needed to square the electric and magnetic fields in the
Hamiltonian.

To find the electric field, we use (3.3), (3.8), and the torsion tensor
$$
{T_{jk}}^m=\Gamma_{jk}^m-\Gamma_{kj}^m=
-{K_{jk}}^m+{K_{kj}}^m\eqno{(4.1)}
$$
and obtain
$$\eqalign{
{\delta\psi\over \delta A^a_i} &= 2 \epsilon^{ijk} B^{\ell a} \left(
\tilde{\nabla}_j {\delta \psi\over \delta \phi^{k\ell }}
+{1\over 2}{T_{jk}}^m  {\delta \psi\over \delta \phi^{m\ell }}\right)
\cr
&\equiv 2 \epsilon^{ijk} B^{\ell a }{\cal D}_j
{\delta \psi\over \delta \phi^{k\ell }} ~,}\eqno{(4.2)}  $$
where
$$
\tilde{\nabla}_j {\delta \psi\over \delta \phi^{k\ell }}\equiv
\partial_j {\delta \psi\over \delta \phi^{k\ell }}-
\Gamma_{jk}^m{\delta \psi\over \delta \phi^{m\ell }}-
\Gamma_{j\ell}^m{\delta \psi\over \delta \phi^{km}}+{1\over 2}\partial_j
(\ln\det G)\, {\delta \psi\over \delta \phi^{k\ell }}\, .
$$
This expression is precisely the covariant derivative of a rank 2
covariant tensor density of weight $-1$.  To see that ${\delta \psi\over
\delta \phi^{k\ell }}$ really is a density of weight $-1$, we note that
in order for $G_{ij}$ to be of weight 0, $B^{ai}$ must be a density of
weight $+1$, from which it follows that $\phi^{ij}$ has weight $+2$, and
thus if $\psi [\phi ]$ is an invariant functional, $\delta \psi/\delta
\phi^{k\ell}$ will have weight $-2+1=-1$.  From (4.2) it then follows
that the electric field is, like the magnetic field, a vector density of
weight $+1$.  It can be checked that this weight assignment to $B^{ia}$,
together with the fact that the Levi-Civita symbol $\epsilon^{ijk}$ has
weight $+1$, leads to consistent tensor densities throughout.

We can now easily write down the electric energy density:
$$
{1\over 2}g^2\delta_{i\bar{\imath}}{\delta\psi^{*} \over \delta A^a_i}
{\delta\psi\over \delta A^a_{\bar{\imath}}} =
2g^2\delta_{i\bar{\imath}}\epsilon^{ijk}\epsilon^{\bar{\imath}\bar{\jmath}
\bar{k}}\phi^{\ell\bar{\ell} }\left({\cal D}_j{\delta \psi^{*}\over
\delta \phi_{k\ell }}\right)\left({\cal D}_{\bar{\jmath}}{\delta
\psi\over \delta \phi_{\bar{k}\bar{\ell}}}\right)\, .\eqno{(4.3)}
$$

If we consider wavefunctionals $\psi [G]$ rather than $\psi [\phi ]$,
then we should rewrite the electric field in terms of derivatives w.r.t.
$G_{ij}$.  This is done by simply taking (4.2) together with the chain
rule giving
$$
{\delta\over \delta \phi^{k\ell}}=
{\delta G_{pq}\over \delta \phi^{k\ell}} {\delta\over \delta G_{pq}}
={1\over 2\det G}\left( G_{pq}G_{k\ell} -2G_{pk}G_{\ell q}\right)
 {\delta\over \delta G_{pq}}\, .\eqno{(4.4)}
$$
Due  to metric compatibility, it is possible to push the covariant
derivative through this prefactor, and we find
$$
{\delta\psi\over \delta A^a_i}={\epsilon^{ijk}B^{\ell a}\over\det G}
\left( G_{pq}G_{m\ell}-2G_{pm}G_{\ell q} \right)
\left(\delta_k^m\tilde{\nabla}_j -{K_{jk}}^m\right)
{\delta\psi\over\delta G_{pq}}\, ,\eqno{(4.5)}
$$
with the {\it caveat} that this covariant derivative is now acting on a
contravariant density of weight $+1$, so that connections and the density
term are changed in sign from (4.2).

We note also that the electric energy density,
$$
{1\over 2}g^2\delta_{i\bar{\imath}} {\delta\psi^{*}\over \delta A^a_i}
{\delta \psi \over \delta
A^a_{\bar{\imath}}}\equiv\delta_{i\bar{\imath}} {\cal
E}^{i\bar{\imath}}\eqno{(4.6)}
$$
is given by the contraction of a fixed Cartesian tensor,
$\delta_{i  \bar{\imath}}$, with a rank 2 contravariant tensor of
$GL(3)$ with
weight 2. Alternatively, one can use (4.5) and form the contraction
$$
\delta_{i \bar{\imath}}\epsilon^{ijk}\epsilon^{\bar{\imath}\bar{\jmath}
\bar{k}} =
\delta^{j
\bar{\jmath}}\delta^{k\bar{k}}-\delta^{j\bar{k}}\delta^{k\bar{\jmath}}
\eqno{(4.7)}
$$
which is another fixed Cartesian tensor which then multiplies a fourth
rank $GL(3)$ tensor of definite symmetry.  One can see from (2.14) that
the $GL(3)$ properties of the magnetic energy density are exactly the
same.  In fact, this $GL(3)$ behavior also holds in the original
variable $A^a_i$ if it is taken to transform as a covariant vector.
Because we are now dealing with the explicit geometric variable
$G_{ij}$, one may hope that the definite $GL(3)$ transformation property
of the Hamiltonian might lead to a group theoretic approach to gauge
field dynamics.

We now wish to give a preliminary discussion about the energy barriers
which appear because we have reexpressed the theory in terms of gauge
invariant variables.  It is clear that the transformation involves both
$G_{ij}$ and its inverse $G^{ij}$, so that there is a singularity when
$\det G=\det B=0$.  Indeed, there are explicit singular factors of
$(1/\det G)$ in (4.5), and more singularities in the connection which
enters both (4.3) and (4.5).  We will be more specific about the nature
of these energy barriers in Section 7, where we restrict to submanifolds
of the function space where we can find explicit expressions for
${K_{ij}}^k$.  However, it is clear that wavefunctionals of finite
energy must vanish in a certain manner for field configurations $B^{ia}$
or $G_{ij}$ whose determinant vanishes somewhere in space.

\goodbreak\bigskip
\noindent{\bf 5.\quad Inversion of the Transformation
$A^a_i\rightarrow B^{ai}\rightarrow G_{ij}$}
\medskip\nobreak

A basic assertion of our approach to $SU(2)$ gauge theory is that any
locally gauge invariant variable, such as the contortion ${K_{ij}}^k$,
can be expressed in terms of the tensor $G_{ij}$.  As we will see, we
must expect these expressions to be non-local, and they seem to be
bi-unique; e.g. there are two configurations ${K_{ij}}^k$ for each
configuration of $G_{ij}$.  It is correct that any functional $\psi
[G_{ij}]$ is gauge invariant, but it is certainly not convenient to
express all gauge-invariant functionals in this form.  One can also
construct gauge-invariant functionals using, e.g., Wilson loops and
Chern-Simons terms.  For this we would like to be able to reconstruct
$B^{ia}$ and $A^a_i$, up to an $SU(2)$ gauge transformation, from
$G_{ij}$.  The forward map $A^a_i\rightarrow B^{ia}\rightarrow G_{ij}$
automatically satisfies the geometrical conditions (3.10, .12, .16), but
we would like to know how big is the image of the space of vector
potentials $A^a_i$ within the space of symmetric tensors $G_{ij}$.  This
is the type of question we discuss in this section.  We will give
reasonable arguments that the situation is favorable but there is more
to be done.

Let us first discuss the reconstruction of $A^a_i$ given $G_{ij}$ and
$\Gamma^k_{ij}$.  This is elementary, but specific to the gauge group
$SU(2)$, which is locally isomorphic to the tangent space group of a
3-manifold.  We consider the quantities
$$\eqalign{
b^{ia}(x)&={1\over |\det B(x)|^{1/2}}\, B^{ia}(x)\cr
{b^{a}}_{i}(x)&= |\det B(x)|^{1/2}\, {B^{a}}_{i}(x)
}\eqno{(5.1)}  $$
which are matrix inverses. For $\det B>0$, we see from (2.11) that
${b^{a}}_{i}$ is an orthonormal frame for $G_{ij}$, and it is well-defined
when $\det B<0$. Thus we start the reconstruction by diagonalizing
$G_{ij}(x)$, writing
$$
G_{ij}(x)=\pm {{\cal R}^a}_i(x)\lambda_a(x){{\cal R}^a}_j(x)\eqno{(5.2)}
$$
where the upper(lower) sign refers to the the case $\det G>0(<0)$, and $
{{\cal R}^a}_i(x)$ is a special orthogonal matrix. The eigenvalues satisfy
$\lambda_a>0$. We then take
$$
{b^a}_i =\pm\sqrt{\lambda_a}{{\cal R}^a}_i~~~~~~{\rm
no~sum~on~}a.\eqno{(5.3)}
$$
Any other ``frame " is related to this by application of an orthogonal
matrix on the left.  The magnetic field associated with the ``metric"
$G_{ij}$ is then defined by
$$
B^{ia}=\sqrt{\det G}\, b^{ia}\, .\eqno{(5.4)}
$$
We then rewrite (3.9) as
$$\eqalign{
\Gamma^i_{jk}&=b^{ia}D_j{b^a}_k\cr
&=b^{ia}(\partial_j{b^a}_k+\epsilon^{abc}A^b_j{b^c}_k)\, ,}\eqno{(5.5)}
$$
which can be rearranged as a ``dreibein postulate"
$$
\partial_j{b^a}_k-\Gamma^{i}_{jk}{b^a}_i+\epsilon^{abc}A^b_j{b^c}_k=0
\eqno{(5.6)}
$$
from   which we can identify the vector potential as the spin
connection, viz.,
$$\eqalign{
\epsilon^{abc}A^c_j&=-\omega_j^{ab}\cr
A^c_j &=
-{1\over 2} \epsilon^{abc} b^{ka}(\partial_j{b^b}_k-
\Gamma^{i}_{jk}{b^b}_i) ~.} \eqno{(5.7)}
$$

{}From the standpoint of the inverse map, it may not be clear that
$B^{ia}$ and $A^a_i$ now defined, respectively, as the frame and spin
connection for the geometry, satisfy the gauge theory relation (2.2).
However, (2.2) is a direct consequence of the Einstein space condition
(3.11).  Contracting this with ${b^a}_\ell b^{kb}$ one finds (for both
signs of $\det B$!),
$$\eqalign{
{R^{ab}}_{ij} &=\partial_i\omega_j^{ab}
-\partial_j\omega_i^{ab}+\omega_i^{ac}
\omega_j^{cb}-\omega_j^{ac}\omega_i^{cb}\cr
&=-\det B({B^a}_i {B^b}_j -{B^a}_j {B^b}_i)\cr
&=-{1\over\det B}\epsilon^{abc}\epsilon_{ijk}B^{kc}\, .}\eqno{(5.8)}
$$
This is equivalent to (2.2) if (5.7) is used to relate $\omega$ and $A$!

Since both $G_{ij}$ and $\Gamma_{ij}^k$ are needed to reconstruct
$A^a_i$, we must ask how to find $\Gamma$, given $G$.  This means that
one must be able to solve (3.12) to find a contortion tensor which
satisfies (3.16).  To see what this involves, we expand out (3.12) by
splitting $\Gamma =\overcirc{\Gamma}-K$, finding, with the help of
(3.16),
$$
\overcirc{R}_{ij}-\overcirc{\nabla}_\ell{K_{ji}}^\ell
-{K_{jm}}^\ell{K_{\ell i}}^m= -2G_{ij}\, . \eqno{(5.9)}
$$
This constitutes 9 equations for the 6 independent components of $K$,
but it turns out that there is a Bianchi identity which imposes 3
relations among these equations, thus one expects that there is a
solution for ${K_{ij}}^k(x)$ for any given configuration $G_{ij}(x)$.

To analyze (5.9) we first insert the representation (3.17) which gives
$$
\overcirc{R}^{j}_{i}-{\epsilon^{j\ell m}\over\sqrt{\det G}}
\overcirc{\nabla}_\ell S_{im}+{\rm sgn}(\det G)
(S^p_iS^j_p-S^p_pS^j_i)=-2\delta^j_i
\eqno{(5.10)}
$$
The first and last terms on the left side are symmetric in $ij$, so the
${\epsilon^{ki}}_j$ contraction of the equation gives the homogeneous
equation
$$
\overcirc{\nabla}_j(S^{kj}-G^{kj}S^p_p)=0\, .\eqno{(5.11)}
$$
while the $\delta^i_j$ contraction is the purely algebraic condition
$$
\overcirc{R} +{\rm sgn}(\det G)
(S^p_iS^i_p-S^p_pS^i_i)= -6\, .\eqno{(5.12)}
$$
It is natural to try to generalize the known Bianchi identity
$$
\overcirc{\nabla}_j (\overcirc{R}^{j}_{i}
- {1\over2}\delta^j_i \overcirc{R})=0\, .\eqno{(5.13)}
$$
by applying $\overcirc{\nabla}_j$ to the difference between (5.10) and
${1\over2}\delta^j_i$ times the trace (5.12). The use of (5.11) helps to
cancel many $(\overcirc{\nabla}S)S$ terms leading to
$$
0=-{\epsilon^{j\ell m}\over\sqrt{\det G}}\overcirc{\nabla}_j
\overcirc{\nabla}_\ell S_{mi}
+{\rm sgn}(\det G)(\overcirc{\nabla}_jS^m_i-\overcirc{\nabla}_iS^m_j
)S^j_m \,
.\eqno{(5.14)} $$
The next step is to compute the product of ${\epsilon_j}^{pq}S^i_q$ with
(5.10). $SSS$ terms miraculously cancel in the resulting expression:
$$
{\epsilon_j}^{pq}\overcirc{R}^j_iS^i_q-{\det G\over\sqrt{\det G}}
(\overcirc{\nabla}^pS^q_i-\overcirc{\nabla}^qS^p_i )
S^i_q=0 \, .\eqno{(5.15)}
$$
This is exactly what is needed to convert the last term in (5.14) into an
algebraic expression which then gives
$$
0=\epsilon^{j\ell m}\overcirc{\nabla}_j
\overcirc{\nabla}_\ell S_{mi}+\epsilon_{ijk}\overcirc{R}^j_mS^{mk} ~.
\eqno{(5.16)}
$$
Finally this reduces to $0=0$ when the Ricci identity and the
representation (3.15) of the curvature tensor are used.  It is possible
to investigate the question of independence of the 9 Einstein equations
directly from (3.12) using the known form \straumann\ of Bianchi
identities with torsion, but we prefer the present determination which
generalizes the familiar Riemannian identity (5.13) by intricate but
straightforward manipulation of torsion terms in (5.10).

\goodbreak\bigskip
\noindent{\bf 6.\quad Sample Geometries}
\medskip\nobreak

In order to develop a better intuition concerning the new spatial
geometry, we consider some particular examples of gauge field geometries
and obtain the metrics $G_{ij}$ and contortions $K_{ijk}$ associated
with them.  In other words, we explore the spatial geometry associated
with some submanifolds of the function space of $SU(2)$ gauge theory.
Among other things, we will find a fringe benefit of the $GL(3)$
invariance.  Namely, we can automatically extend our solutions to orbits
of the group of diffeomorphisms.

We first explore the case where the magnetic field $B^{ia}$ is constant
in some gauge in a finite region $V\subset {\IR}^3$, and the potential
$A^a_i$ is also constant there (we exclude abelian configurations with
linear potentials).  Then
$$
B^{ia}={1\over2}\epsilon^{abc}\epsilon^{ijk}A^b_jA^c_k\, ,\eqno{(6.1)}
$$
which implies $\det B=(\det A)^2\ge 0$. Equation (6.1) can be inverted to
find two solutions for $A$, namely
$$
A^a_i=\pm \sqrt{\det B}{B^a}_i\, .\eqno{(6.2)}
$$
Thinking in terms of the forward map, we compute from (2.11) and (6.2)
$$
G_{ij}=A^a_iA^a_j\eqno{(6.3)}
$$
which is an arbitrary positive constant matrix (see (2.15)).

To obtain the connection, which is pure contortion for constant fields,
we use the definition (3.9), and obtain
$$\eqalign{
{K_{ij}}^k&={B^a}_jD_iB^{ka}\cr
&=\pm {{\epsilon_{ij}}^k\over\sqrt{\det G}}   }\eqno{(6.4)}
$$
where the last line requires (6.2,.3) and some calculation using
properties of the inverse of $3\times 3$ matrices.  Comparing with
(3.17) we see that the tensor $S^{ni}=\pm G^{ni}$.

One can also derive (6.4) directly from the Einstein condition (5.9). For
constant fields $\overcirc{R}_{ij}$ and $\overcirc{\nabla}_\ell
{K_{ij}}^\ell$ vanish, and (5.9) becomes algebraic, viz.,
$$
{K_{jm}}^\ell {K_{\ell i}}^m=-2G_{ij}\, .\eqno{(6.5)}
$$
It is easy to show that the only solutions are given by (6.4).  Thus the
contortion tensor is totally antisymmetric, and it is minimally
constructed from the metric $G_{ij}$, hence covariantly constant,
$$
\overcirc{\nabla}_\ell {K_{ij}}^\ell =0\, .\eqno{(6.6)}
$$
Of course, the Christoffel geometry is flat, i.e.,
$\overcirc{R}_{ij}=0$.

We may now generalize this solution by introducing arbitrary ``coordinate
functions" $y^\alpha (x^i)$, with
$$
\det {\partial y^\alpha\over\partial x^i}\neq 0
$$
and considering the new metric
$$
G'_{\alpha\beta}(y(x))=G_{ij}{\partial x^i\over\partial y^\alpha}
{\partial x^j\over\partial y^\beta}\eqno{(6.7)}
$$
which  is ``diffeomorphic" to $G_{ij}$. Then the transformed version of
(5.9) will have the solution
$$\eqalign{
{K'_{\alpha\beta}}^\gamma &=
\pm {{\epsilon_{\alpha\beta}}^\gamma\over\sqrt{\det G'}}\cr
\overcirc{R}'_{\alpha\beta}&=0\, ,}
\eqno{(6.8)}
$$
and one can then reconstruct gauge fields $B'^{\alpha a}(y(x))$ and
$A'^a_{\alpha }(y(x))$ which are ``diffeomorphic" to  $B^{ia}(y(x))$ and
$A^a_{i}(y(x))$. The new fields are not constant, and they are related by
$$
B'^{i\alpha}(y(x))={1\over2}\epsilon^{abc}\epsilon^{\alpha\beta\gamma}
A'^b_\beta A'^c_\gamma ~.
\eqno{(6.9)}
$$
It is in this way that a correspondence between gauge field configurations
and geometries on subspaces of the function space of the theory can be
extended by action of the group of diffeomorphisms.

There is one other important implication of this sample geometry.
Although we do not know the general solution of (5.9), it presumably
must reduce to (6.4), whenever $\overcirc{R}_{ij}(x)$ vanishes in any
finite subset of ${\IR}^3$.  It therefore seems correct to infer that in
the general case, there are always two solutions ${K_{ij}}^k$ for a
given metric $G_{ij}$.

The second  class of gauge field configurations to be studied are those
where, in some gauge, the potential takes the form
$$
A_i^a(x)=\lambda {\cal A}_i^a(x)\eqno{(6.10)}
$$
where $\lambda$ is any real number and ${\cal A}_i^a(x)$ is a pure gauge, i.e.,
$$
\partial_i {\cal A}_j^a(x)-\partial_j {\cal A}_i^a(x) +
\epsilon^{abc}{\cal A}_i^b(x) {\cal A}_j^c(x)=0\, .\eqno{(6.11)}
$$
The magnetic field is then
$$\eqalign{
B^{ia}&=\lambda (\lambda -1){1\over2}\epsilon^{ijk}
\epsilon^{abc}{\cal A}_j^b {\cal A}_k^c\cr
&=\lambda (\lambda -1){\cal A}^{ia}\,\det {\cal A}\, ,}
\eqno{(6.12)}
$$
where $\det {\cal A}=\det {\cal A}_i^a$ and ${\cal A}^{ia}$ is the
matrix inverse of ${\cal A}_i^a$.  These gauge fields were first studied
in \deser\ as an example of the Wu-Yang ambiguity, in which different
potentials, for $\lambda$ and $1-\lambda$, give the same magnetic field.

The gauge-invariant ``metric" variable $G_{ij}$ is given by
$$
G_{ij}(x)=\lambda (\lambda -1){\cal A}_i^a(x) {\cal A}_j^a(x) ~.
\eqno{(6.13)}
$$
One sees that $\det B=\det G=[\lambda (\lambda -1)]^3 (\det {\cal A})^2$, so
we have $\det G<0$ for $0<\lambda <1$, and $\det G>0$ otherwise.

To understand this geometry we use the representation
$$
-{i\over2}\tau^a {\cal A}_i^a=U^{-1}\partial_i U\eqno{(6.14)}
$$
where $U(x)$ is an arbitrary $SU(2)$ matrix. Then (6.11) is satisfied, and
one finds
$$
G_{ij} = 2 \lambda (\lambda -1){\rm ~Tr~} \partial_i U\partial_jU^{-1} ~.
\eqno{(6.15)}
$$
One can then write
$$\eqalign{
U(x)&=\alpha_4(x) +i\vec{\tau}\cdot\vec{\alpha}(x)\cr
&\sum_{s=1}^4\,\alpha_s^2=1\, ,
}\eqno{(6.16)}
$$
and obtain
$$
G_{ij}(x)=4\lambda (\lambda -1)
\sum_{s=1}^4 (\partial_i\alpha_s) (\partial_j\alpha_s) ~.
\eqno{(6.17)}
$$
Since
$$\eqalign{
ds^2&\equiv G_{ij}dx^idx^j\cr
&=4 \lambda (\lambda -1)\sum_{s=1}^4 d\alpha_s d\alpha_s }\eqno{(6.18)}
$$
with $\alpha_s$ a unit $4$-vector, one sees that $G_{ij}$ is
proportional to a metric on the round $3$-sphere.  Globally there may be
multiple coverings of the sphere by the map $\alpha_s(x)$ from
${\IR}^3$.  In this case a diffeomorphism is implemented by
$\alpha_s'(y(x))=\alpha_s(x)$ and ${\cal A}'^a_\alpha (y(x))= (\partial
x^i/\partial y^\alpha )\, {\cal A}^a_i(x)$.  This leaves us within the
initially defined class of potentials.

To compute the connection one uses the definition (3.9),
separated into two terms:
$$
{B^{ka}\over\sqrt{\det G}}\Gamma^i_{jk}=-\left[ \partial_j\left(
{B^{ia}\over\sqrt{\det G}}\right)+\epsilon^{abc}A^b_j
{B^{ic}\over\sqrt{\det G}}\right]\, .\eqno{(6.19)}
$$
The second term is similar to that of the constant case,
$$
-\epsilon^{abc}A^b_j {B^{ic}\over\sqrt{\det G}}=\pm \left(
{B^{ka}\over\sqrt{\det G}}\right)
{\lambda{\epsilon^i}_{jk}\over |\lambda (\lambda -1)|^{1/2}\sqrt{\det G} } ~.
\eqno{(6.20)}
$$
To treat the first term we use (6.12) and obtain
$$
-\partial_j\left(
{B^{ia}\over\sqrt{\det G}}\right)={B^{ka}\over\sqrt{\det G}}
{\cal A}^{ib}\partial_j{\cal A}_k^b\, .\eqno{(6.21)}
$$

The  last factor is split onto symmetric and antisymmetric terms in
$jk$. One uses (6.11) for the latter and (6.11) plus the definition of
$\overcirc{\Gamma}^{i}_{jk}$ in (3.13) for the former. The result is
$$
-\partial_j\left(
{B^{ia}\over\sqrt{\det G}}\right)={B^{ka}\over\sqrt{\det G}}
\left[ \overcirc{\Gamma}^{i}_{jk}\mp {1\over2}
{{\epsilon^i}_{jk}\over|\lambda (\lambda -1)|^{1/2}\sqrt{\det
G}}\right] ~.\eqno{(6.22)}
$$

The contortion tensor can be identified from (6.20) and (6.22)  as
$$
{K_{jk}}^i=\mp {\lambda-{1\over2}\over |\lambda (\lambda -1)|^{1/2}}
{{\epsilon^i}_{jk}\over \sqrt{\det G}}\, .\eqno{(6.23)}
$$
The sign in (6.23) is the product
$\mp=-{\rm sgn}[\lambda (\lambda -1)]\, {\rm sgn}\det {\cal A}$.
So the contortion is again the minimal totally antisymmetric structure.

The two Wu-Yang related potentials $\lambda {\cal A}$ and $(1-\lambda
){\cal A}$ give contortion tensors of opposite signs, as in (6.23),
although the interpretation of the signs is best stated in terms of the
inverse map.  Suppose one is given a metric of the form (6.18) and a
pair of contortion tensors of the form
$$
{K_{jk}}^i=\mp {|\lambda-{1\over2}|\over |\lambda (\lambda -1)|^{1/2}}
{{\epsilon^i}_{jk}\over \sqrt{\det G}}\, .\eqno{(6.24)}
$$
Then from $G$ and $K$ with the upper sign, one will reconstruct via
(5.4) and (5.7) a potential $A^a_i(x)$ which is in general not
proportional to a pure gauge, but is the gauge transform of either
$\lambda {\cal A}$ or $(1-\lambda ){\cal A}$.  Then for $G$ and $K$ with
lower sign, the reconstruction leads to another potential which is gauge
equivalent to the Wu-Yang related form $(1-\lambda ){\cal A}$ or
$\lambda {\cal A}$, respectively.

The 3-sphere is an Einstein space, and one can compute by the standard
method of the Cartan structure equations that the metric (6.17)
satisfies
$$
\overcirc{R}_{ij}={1\over2\lambda(\lambda -1)}G_{ij}\, .\eqno{(6.25)}
$$

This suggests that we investigate special solutions of the geometrical
equations (5.9) in which $G_{ij}$ is an Einstein space in the Riemannian
sense and ${K_{jk}}^i$ is minimal, viz.,
$$\eqalign{
\overcirc{R}_{ij}&=\Lambda G_{ij}\cr
{K_{ij}}^k&= c{{\epsilon^k}_{ij}\over \sqrt{\det G}}\, .}\eqno{(6.26)}
$$
Then (5.9) is satisfied if the parameters are related by
$$
c^2={\rm sgn}(\det G)(\Lambda +1)\, .\eqno{(6.27)}
$$
If $\Lambda >0$, then (6.26) is satisfied by a positive definite
$G_{ij}$ which is locally a round metric on $S^3$.  Since
$\overcirc{R}_{ij}(G)=\overcirc{R}_{ij}(-G)$, we see that for $\Lambda
<0$, (6.26) is satisfied by a negative-definite $G_{ij}$ which is the
negative of a round $S^3$ metric.  In this case the torsion is real only
if $\Lambda <-1$.  Both ranges $\Lambda >0$ and $\Lambda <-1$ are
covered if we take $\Lambda =1/(4\lambda (\lambda -1))$ as in (6.25),
and it is easy to see that the magnitude of the contortion $c$ agrees
with (6.23).  Thus the solutions of (6.26) we are now discussing give
Deser-Wilczek potentials under the inverse map.  We believe that they
exhaust this class of potentials, although we are uncertain of global
issues such as multiple coverings of $S^3$.

If $\Lambda <0$, then there are certainly positive definite metrics
$G_{ij}$ which are solutions of (6.26).  These should also correspond to
gauge field configurations, at least in the range $-1<\Lambda <0$ in
which the torsion is real.

We also suggest that it may be useful to investigate solutions of (5.9)
restricted by the condition of spherical symmetry.  This should lead to
a system of ordinary differential equations for 4 radial functions, 2
contained in the most general {\it ansatz} for a spherically symmetric
$G_{ij}$ and 2 more for $S^{ij}$.  The corresponding gauge field
configurations should include monopole solutions of Yang-Mills-Higgs
systems with all but spatial potentials $A^a_i$ and $B^{ia}$ discarded.

\goodbreak\bigskip
\noindent{\bf 7.\quad  Energy Barriers}
\medskip\nobreak

We  now write the energy density in a form suitable for our discussion.
It is useful to define the quantities

$$\eqalign{
V^{impq}&=\epsilon^{ijk} \left( \delta_k^m
\tilde{\nabla}_j -{K_{jk}}^m\right){\delta\psi\over\delta G_{pq}}\cr
Q^{im}&=G_{pq}(V^{impq}-2V^{iqmp})\, .}\eqno{(7.1)}
$$
Then  by straightforward index-shuffling one can reexpress the energy
density (4.6) in the form
$$
\delta_{i\bar{\imath}} {\cal E}^{i\bar{\imath}}=
{1\over 2}g^2\delta_{i\bar{\imath}}
{G_{m\bar{m}}\over\det G} Q^{*\bar{\imath} \bar{m}} Q^{im} ~.
\eqno{(7.2)}
$$
Separating out the torsion from the derivative $\tilde{\nabla}$ in (7.1),
one can obtain
$$\eqalign{
Q^{im}=G_{pq}& \left( \epsilon^{ijm} \tilde{\overcirc{\nabla}}_j
{\delta\psi\over\delta G_{pq}}-2 \epsilon^{ijq} \tilde{\overcirc{\nabla}}_j
{\delta\psi\over\delta G_{pm}} \right)\cr & +\epsilon^{ijk}
\left( {K_{jk}}^m
G_{pq}{\delta\psi\over\delta G_{pq}}-2 {K_{jn}}^m
G_{pk}{\delta\psi\over\delta G_{pn}}\right) }
\eqno{(7.3)}
$$
where $\tilde{\overcirc{\nabla}}_j$ is a torsion-free covariant derivative
with density term.

We now specialize to the cases discussed in Section 6 where the contortion
takes the minimal form (6.26). The torsion term in (7.3) then simplifies,
leading to
$$\eqalign{
Q^{im}=G_{pq}& \left( \epsilon^{ijm} \tilde{\overcirc{\nabla}}_j
{\delta\psi\over\delta G_{pq}}-2 \epsilon^{ijq}
\tilde{\overcirc{\nabla}}_j {\delta\psi\over\delta G_{pm}} \right)\cr
& +2c{\rm ~sgn}(\det G)\sqrt{\det G}{\delta\psi\over\delta G_{im}}\, .}
\eqno{(7.4)}
$$
When this is substituted in (7.2), one finds an expression for the
energy density which can be seen to contain terms with spatial
derivatives which are singular when $\det G\rightarrow 0$ at any point
in space and terms without spatial derivatives which are regular.
Positive definiteness implies that the singularity does not cancel in
any simple way.  Instead the gauge invariant state functional $\psi [G]$
must be constrained in a complicated way near the barrier in order to
have finite total energy.  The energy barriers are certainly present for
general configurations $G_{ij}$ for which (7.3) must be used.  They are
simply clearer when the restriction is made to one of the flat or
constant curvature configurations of Section 6 because the contortion
tensor is explicitly known.  One can also consider the case where a
field configuration $G_{ij}$ reduces to a constant curvature or flat
metric on some finite subset $V\subset {\IR}^3$.  The form (7.4) is then
valid for $x\in V$, but the functional $\psi [G]$ and its derivatives
depend on the behavior of $G_{ij}$ throughout space.

There are similar energy barriers at $\det\phi =0$, if the energy
density is expressed in terms of the $\phi^{ij}$ variables as in (4.3),
and they are known to be present in the electric formulation \gold\ of
nonabelian gauge theories.  We believe that they are a general feature
of any formulation of the theory in gauge invariant variables, because
the transformation to these variables is nonlinear and inevitably has
singular points.  The perturbative wave functional is not of the form
$\psi [G]$, but because it peaks at zero magnetic field where $\det
B=0$, we would expect significant non-perturbative corrections.

Finally we would like to point out that for fields which are constant
throughout space, spatial derivative terms in (7.4) can be dropped, and
(6.4) tells us that $c=\pm 1$.  The electric energy density then becomes
simply
$$
\delta_{i\bar{\imath}} {\cal E}^{i\bar{\imath}}=
{1\over 2}g^2\delta_{i\bar{\imath}} G_{m\bar{m}}
{\delta\psi^{*}\over\delta G_{\bar{\imath}\bar{m}}}
{\delta\psi\over\delta G_{im}} ~.
\eqno{(7.5)}
$$
Using (6.3) we see that this is equal to
$$
\delta_{i\bar{\imath}} {\cal E}^{i\bar{\imath}}=
{1\over 2}g^2\delta_{i\bar{\imath}} {\delta\psi^{*}\over\delta
A^a_i}{\delta\psi\over\delta A^a_i } ~.
\eqno{(7.6)}
$$
It is no surprise that this coincides with the electric part of the
Hamiltonian used in \luscher ,\baal\ to describe constant modes of the
$SU(2)$ gauge field in a box.

\goodbreak\bigskip
\noindent{\bf 8.\quad  Open Questions}
\medskip\nobreak

The spatial geometry found in the gauge-invariant configuration space of
$SU(2)$ gauge theory has not been recognized previously.  It is a rather
simple geometry, and a natural question is the generalization to larger
gauge groups, notably the true color group $SU(3)$.  Here the situation
is that more independent variables are required to describe the
gauge-invariant degrees of freedom of the magnetic field $B^{ia}$.  In
$SU(3)$ one can take \ken\ the 6 components of
$\phi^{ij}=B^{ia}B^{ja}$and the 10 components of the symmetric tensor
density formed with the $d$-symbol, namely $d^{abc}B^{ia}B^{jb}B^{kc}$.
The rectangular matrix $B^{ia}$ has many ``right-inverses" and in
general no ``left-inverse".  Partly because of this, careful navigation
will be required to find the underlying geometry, but we are optimistic
that it can be found.

It is far from clear that the geometry will be helpful in understanding
the dynamics of the $SU(2)$ gauge theory.  Indeed there are several
problems to be overcome if $G_{ij}$ is to be viewed as the fundamental
variable.  First, one must be able to solve (5.9) for the contortion to
obtain an explicit expression for the Hamiltonian.  Second one must
represent the functional Jacobian $\delta B^{ia}(x)/\delta A^b_j(y)$
which involves the determinant of a differential operator with singular
symbol.  Third one must deal with the fact that a gauge invariant
formulation of the theory, whether magnetic or electric \gold , is
essentially nonperturbative.  There are $1/g$ terms which remain in $H$
after rescaling the fields either by an integral or fractional \luscher\
power of $g$.  In the past geometry has provided strong impetus for
physicists, and we hope that it will stimulate new approaches to the
important problem of gauge field dynamics.\bigskip

\goodbreak\bigskip
\noindent{\bf Appendix.\quad Tranformations of the Measure}
\medskip\nobreak

As mentioned briefly in Sec. 8 a complete change of variables
$A\rightarrow B\rightarrow G$ requires consideration of the Jacobian
determinant.  This will enter in the measure of the transformed
functional integral over $G_{ij}(x)$, and terms from the measure will be
generated when a functional integration by parts is performed in Eq.
(7.2) to put the Hamiltonian in the more standard second derivative
form.

The essential part of the measure is the functional determinant of the
transformation operator from $A$ to $B$, namely
$$M^{IJ}(x,y)\equiv
M^{ia,jb}(x,y)\equiv {\delta B^{ia}(x)\over \delta A^b_j(y) } = \epsilon
^{imj} D_m{}^{ab} \delta^{(3)}(x-y) \eqno{(A.1)}$$
This operator is
gauge covariant, but a manifestly gauge invariant form is useful.  This
can be achieved by conjugating with the ultralocal algebraic operator
${\IB_{jb}}^{k\ell}(x,y)\equiv\delta_j^kB^{\ell b}(x)\delta^{(3)}
(x-y)$. Namely one convolutes ${M_{ia}}^{jb}(x,y)$ with
this operator from
the right and with its inverse from the left.  This gives the new
invariant differential operator
$$\eqalign{
{N_{mn}}^{k\ell}(x,y)&\equiv{\IB^{-1}_{mn}}^{ia}(x,u)\cdot
{M_{ia}}^{jb}(u,v)\cdot {\IB_{jb}}^{k\ell}(v,y)\cr
&={\epsilon_m}^{ik}(\delta_n^\ell\partial_i-\Gamma'^{\ell}
_{in})\delta^{(3)} (x-y)\, , }
\eqno{(A.2)}$$
where (3.2) has been used.  One nows sees explicitly from
(4.2) that $N$ involves the same covariant derivative that occurs in the
transformed electric field.  Formally $M$ and $N$ have the same
determinant,
so we see that the measure has been brought to geometric form.

The remaining terms
in the functional measure are ultralocal factors from the transformation
$\delta G/\delta B$ and the integration over the time-independent
$SU(2)$ gauge group at each point $x$ of $\IR^3$. We simply state the
final result:
$$
[dG_{ij}]\, {\det G\over\det N}\, .\eqno{(A.3)}
$$

Functional determinants in general require ultraviolet regularization,
and this is difficult in the present case where the highest derivative
term of $N$ has zero modes.  We leave such problems for future
investigation because the measure is not of direct concern in this
paper.

\listrefs

\end